\documentclass[twocolumn, prl, preprintnumbers,amsmath,amssymb]{revtex4}

\usepackage{graphicx}% Include figure files
\usepackage{dcolumn}% Align table columns on decimal point
\usepackage{bm}% bold math

\begin{document}

\preprint{}

\title{Bias-controlled sensitivity of ferromagnet/semiconductor electrical spin detectors}

\author{S. A. Crooker$^{1*}$, E. Garlid$^2$, A. N. Chantis$^1$, D. L. Smith$^1$, K. S. M. Reddy$^3$, Q. Hu$^2$, T. Kondo$^{2}$,
C. J. Palmstr{\o}m$^{3,4}$, P. A. Crowell$^2$}

\affiliation{$^1$Los Alamos National Laboratory, Los Alamos, NM
87545}

\affiliation{$^2$School of Physics and Astronomy, University of
Minnesota, Minneapolis, MN 55455}

\affiliation{$^3$Dept. of Chemical Engineering \& Materials Science,
University of Minnesota}

\affiliation{$^4$Depts. of Electrical \& Computer Engineering and
Materials, University of California, Santa Barbara, CA 93106}

%\affiliation{$^*$crooker@lanl.gov}

\date{\today}
\begin{abstract}
Using Fe/GaAs Schottky tunnel barriers as electrical spin detectors,
we show that the magnitude and sign of their spin-detection
sensitivities can be widely tuned with the voltage bias applied
across the Fe/GaAs interface. Experiments and theory establish that
this tunability derives not just simply from the bias dependence of
the tunneling conductances $G_{\uparrow,\downarrow}$ (a property of
the interface), but also from the bias dependence of electric fields
in the semiconductor which can dramatically enhance or suppress
spin-detection sensitivities. Electrons in GaAs with fixed
polarization can therefore be made to induce either positive or
negative voltage changes at spin detectors, and some detector
sensitivities can be enhanced over ten-fold compared to the usual
case of zero-bias spin detection.
\end{abstract}

%\pacs{72.25.Dc, 72.25.Mk, 72.25.Hg, 87.75.-d}
\maketitle
%---------------------------------------------------------------
An all-electrical scheme for efficient detection of spins is an
essential component of most ``semiconductor spintronic" device
proposals \cite{Zutic, Jansen}, and recent demonstrations of
electrical spin injection and detection in GaAs \cite{Johnson2,
Bhatta, LouNP}, silicon \cite{Huang, vantErve}, and graphene
\cite{Tombros} have motivated a desire to understand how bulk and
interfacial spin transport contribute to device performance.
However, while the need for optimizing spin injection is widely
recognized, the interplay of mechanisms that govern electrical spin
\emph{detection} have only begun to be explored \cite{Jansen,
Bhatta, Fert}.

Recent progress on spin-transport devices with semiconductor
channels has relied on spin-dependent electron tunneling through
barriers formed at ferromagnet-semiconductor interfaces
\cite{MJcomment,Rashba,Smith}. However, tunneling typically
introduces a strong and often unpredictable dependence of the
tunneling current polarization, $P_j$, on interface voltage bias.
For example, tunneling magnetoresistances through vertical
Fe/GaAs/Fe trilayers were found to invert sign when one interface
was epitaxial \cite{Moser}. In lateral Fe/GaAs devices, $P_j$ varied
markedly with source electrode bias, unexpectedly inverting sign
under forward bias in some structures and under reverse bias in
others \cite{LouNP}. Accordingly, several recent theories have
attempted to understand spin tunneling through
ferromagnet-semiconductor interfaces \cite{DeryPRL, ChantisPRL,
Fert}. In addressing the relevance of these models to the
equally-important problem of spin detection, there is a critical and
largely unexplored need to separate interfacial tunneling effects
from the effects of spin transport due to electric fields
\cite{YuFlatte, Schmidt} in the semiconductor.

Here we use lateral Fe/GaAs structures to study the influence of a
detector interface bias ($V_d$) on the sensitivity of electrical
spin detection, which we define as the voltage change $\Delta V_d$
induced by an additional, remotely-injected spin polarization.
Crucially, we show that spin-detection sensitivities cannot be
understood from the detector interface's tunneling properties
($P_j$) alone. Instead, both experiments and theory show that
$\Delta V_d$ can be \emph{significantly} enhanced or suppressed
compared to $P_j$ -- in a controlled and predictable fashion -- due
to the resulting electric fields in the semiconductor, which modify
the spin densities ($n_{\uparrow,\downarrow}$) and their gradients
at biased Fe/GaAs detector interfaces. These new phenomena cannot
occur in conventional non-local studies that use spin detectors
operating at zero bias, nor are they expected in all-metal devices
having large channel conductivities and therefore negligible
electric fields.

\begin{figure}[tbp]
\includegraphics[width=.45\textwidth]{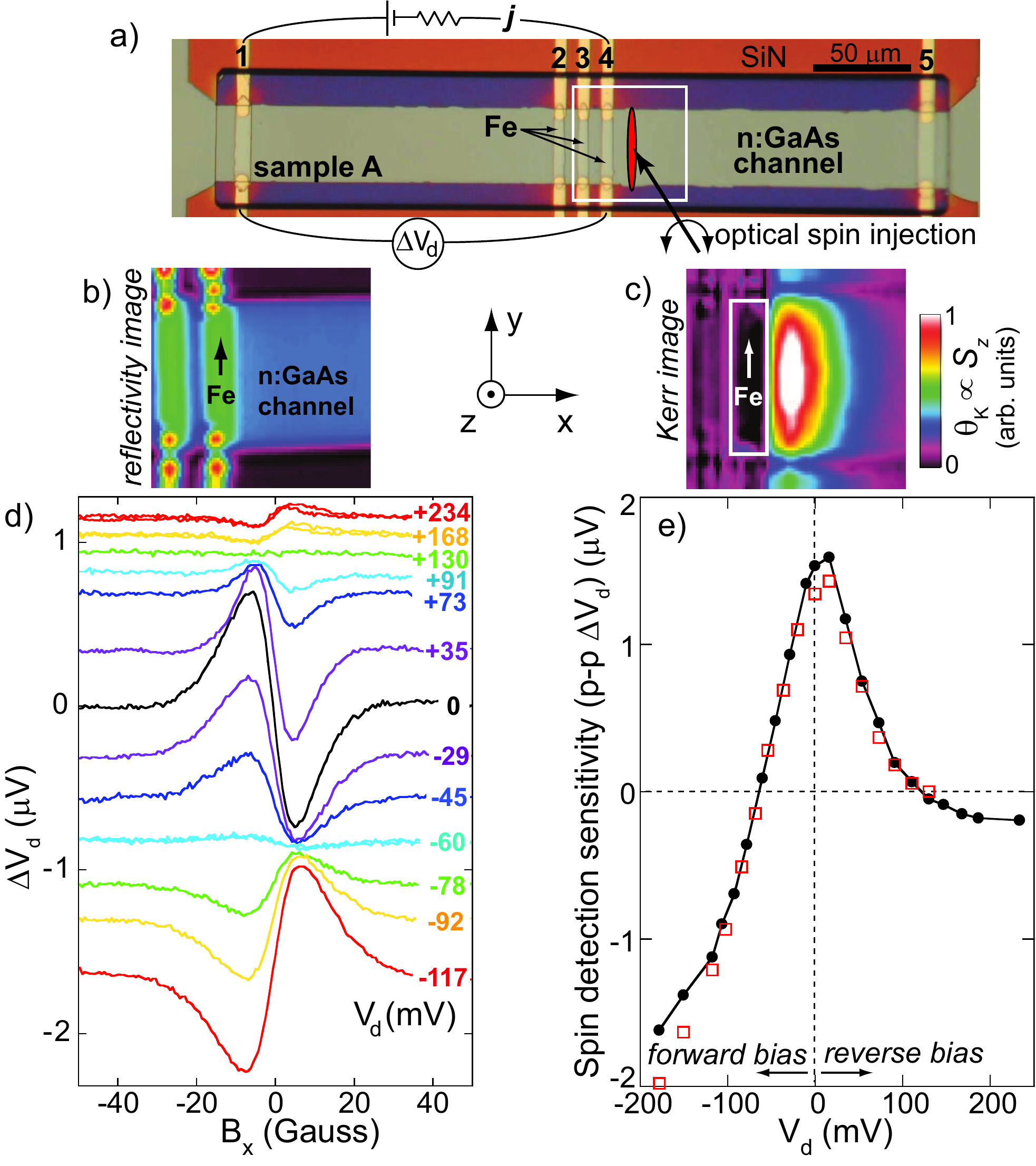}
\caption{(Color online) a) Spin transport device A and the spin detection
experiment. Current \emph{j} establishes a voltage $V_d$ across a
Fe/GaAs detector interface (contact 4). Changes ($\Delta V_d$) due
to a remote, optically-injected spin polarization reveal the
detector's spin-sensitivity. b) A 70$\times$70 $\mu$m reflectivity
image [white square in (a)], showing device features. c) A
Kerr-rotation image of the optically-injected polarization
($B_x$=0). d) Raw spin-detection data at 10~K, $\Delta V_d(B_x)$, at
various $V_{d}$. 4~K and 20~K data are very similar. e) The
corresponding spin-detection sensitivity versus $V_{d}$ (black
points). Squares show similar $\Delta V_d$ measured for injection
\emph{in} the current path ($j$ flowing between contacts 4 and 5).}
\label{fig1}
\end{figure}

Fig. 1(a) shows a typical device. Epitaxial Fe/GaAs structures are
processed into lateral spin-transport devices having five
10$\times$50 $\mu$m Fe contacts (\#1-5) on a 2.5 $\mu$m thick
\emph{n}-type GaAs channel \cite{LouNP}. A Schottky tunnel barrier
is formed at the interface between Fe and a highly-doped $n^+$
interfacial layer of GaAs \cite{Hanbicki}. Devices from two wafers
are discussed (denoted A and B, with $n$=3.5 and
2.0$\times$10$^{16}$ cm$^{-3}$); these represent structures for
which $P_j$ inverts under finite forward and reverse source bias,
respectively \cite{LouNP}.

To measure the spin-detection sensitivity of these Fe/GaAs
electrodes in a background-free manner, we employ an optical pumping
technique. First, a constant current $j$ establishes a voltage drop
$V_d$ across a Fe/GaAs detector interface (contact \#4 in Fig. 1).
From \cite{Smith},
\begin{equation}V_d = \frac{j}{4}\left[\left( \frac{1}{G_\uparrow }
+ \frac{1}{G_ \downarrow} \right) + \left( \frac{1}{G_ \uparrow }
-\frac{1}{G_ \downarrow}\right)P_j \right],
\end{equation}
where $P_j =(j_\uparrow - j_\downarrow)/j$ is the tunneling current
polarization, and $j_{\uparrow, \downarrow}$ and $G_{\uparrow,
\downarrow}$ are the majority- and minority-spin tunneling currents
and conductances. A weak, circularly-polarized, 1.58 eV pump laser
is then focused to a stripe 10-20 $\mu$m away from contact 4,
injecting a small constant spin polarization in the channel. This
additional polarization, oriented initially along $\pm \hat{z}$,
drifts and diffuses laterally and is tipped into the minority or
majority spin direction with respect to the Fe magnetization
\textbf{M} ($\parallel$$\pm \hat{y}$) by small applied fields $\pm
B_x$. At the Fe/GaAs detector interface, this polarization modifies
the chemical potentials $\mu_{\uparrow , \downarrow}$ which (at
constant $j$) necessarily modifies $P_j$ and therefore changes $V_d$
by an amount $\Delta V_d$ that is a direct measure of the detector's
spin sensitivity. The dependence of $\Delta V_d$ on $V_d$ derives
both from the bias dependence of $G_{\uparrow\downarrow}$ (a
property of the interface), \emph{and} from the optically-induced
changes to $P_j$ (which depend on electric fields in the GaAs, as
detailed later).

To explicitly measure spin-dependent changes, the laser is modulated
from right- to left-circular polarization at 50 kHz, and $\Delta
V_d$ at this frequency is measured between the detector and a
distant reference contact. Importantly, this approach avoids
magnetic dichroism and hot electron artifacts because spins are
injected into the channel (rather than through the Fe contact),
where they cool before diffusing. Using
scanning Kerr-rotation microscopy \cite{Stephens, CrookerScience,
FurisNJP, Kotissek}, Figs. 1(b,c) show images of the reflectivity, and of the
optically-injected polarization diffusing to the right and also to
the left towards the spin detector. Using low laser power ($\sim$10
$\mu$W), perturbations to $\mu_{\uparrow , \downarrow}$ are
intentionally kept small, of order 1 $\mu$V.

Fig. 1(d) shows $\Delta V_d$ versus $B_x$ at different biases
$V_{d}$ across the Fe/GaAs detector. Effectively, we electrically
detect the $\hat{y}$ component of diffusing and precessing spins
that are optically oriented initially along $\pm \hat{z}$. Thus, the
data show antisymmetric ``local Hanle" lineshapes that depend on
spin lifetimes, diffusion constants, and source/detector separation
\cite{CrookerScience, FurisNJP}. Maximum and minimum $\Delta V_d$
occur at $\pm B_x$ for which the injected spins precess on average
$\pm 90^\circ$ ($\parallel$$\pm$\textbf{M}) upon reaching the
detector. Larger $B_x$ causes rapid ensemble dephasing.

\begin{figure}[tbp]
\includegraphics[width=.45\textwidth]{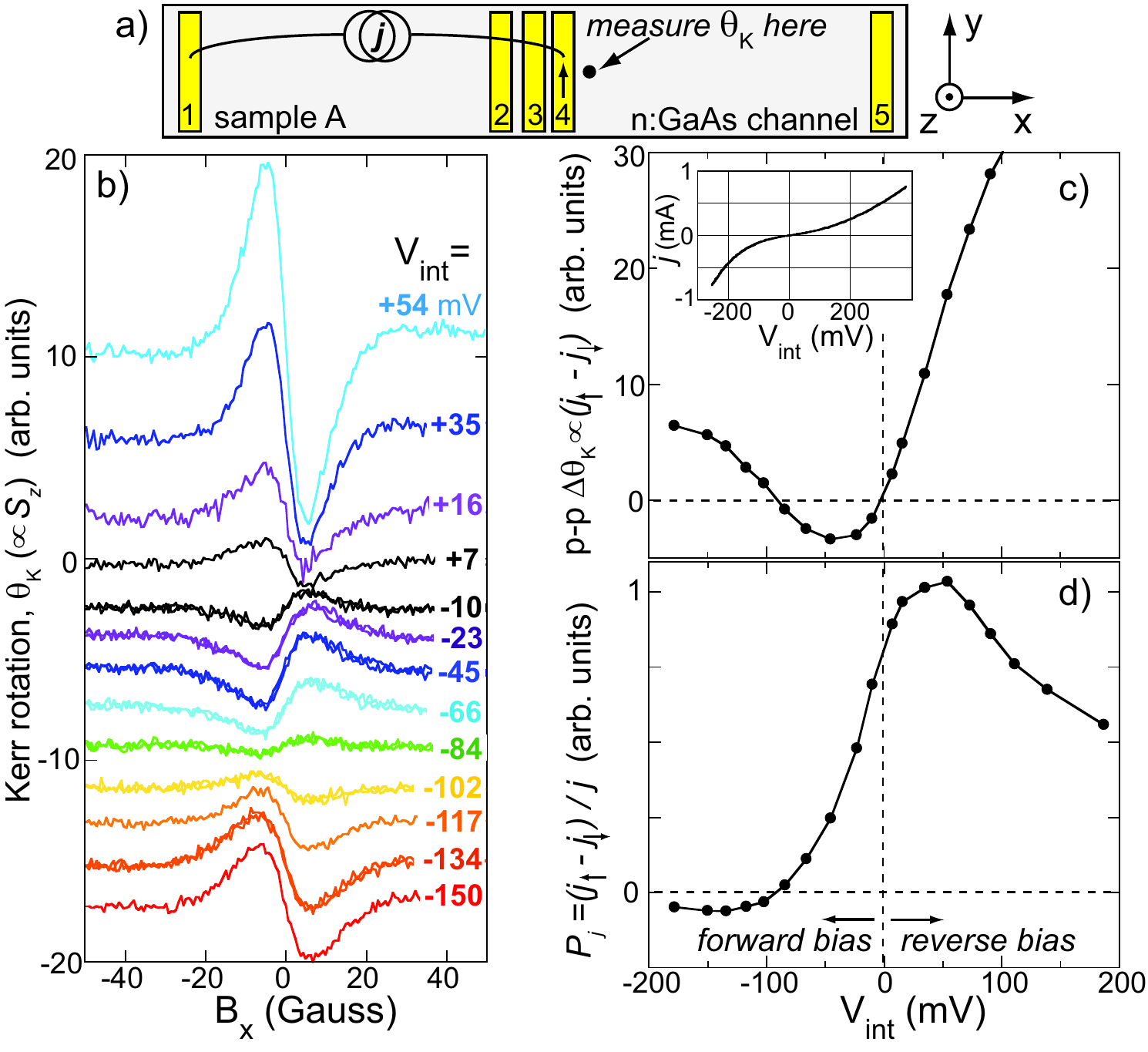}
\caption{(Color online) a) Measuring $P_j$, the tunneling current polarization at
contact 4 via spin \emph{injection} at 10~K. Current $j$ establishes
a voltage bias $V_{int}$ across the Fe/GaAs interface. b)
Electrically-injected spins are measured via the Hanle-Kerr effect
($\theta_K$ vs. $B_x$), at different $V_{int}$ (curves offset). c)
The peak-to-peak Hanle amplitude, $\Delta \theta_K \propto
j_\uparrow - j_\downarrow$, versus $V_{int}$. Inset: $j(V_{int})$
for contact 4. d) Relative polarization, $P_j \propto (j_\uparrow -
j_\downarrow)/j$, versus $V_{int}$.} \label{fig2}
\end{figure}

The peak-to-peak amplitudes of these curves [Fig. 1(e)] therefore
provide a relative measure of the electrical spin-detection
sensitivity that depends only on the spin-dependent voltages induced
by the remotely-injected spins. In sample A, $\Delta V_d$ is clearly
tunable with detector bias. It is large and positive near $V_d$=0,
where the remotely-injected spins induce a positive $\Delta V_d$
($\simeq$+1.5 $\mu$V) at the detection electrode. With increasing
forward or reverse detector bias, however, $\Delta V_d$ decreases
and inverts sign -- now, these same spins induce a \emph{negative}
voltage change. Electrons in GaAs with fixed spin can therefore be
made to generate positive \emph{or} negative voltage changes at
these Fe/GaAs spin detectors by tuning $V_d$ by a few tens of mV,
offering new routes (besides manipulating electron spins or contact
magnetizations) to tune and switch spin-dependent electrical
signals.

To understand this tunable spin detection sensitivity, it is
essential to independently measure $P_j$, the polarization of the
tunneling current that flows across this biased Fe/GaAs interface.
For this we measure the electrical spin \emph{injection} efficiency
of this electrode using Kerr-rotation methods, as demonstrated
previously \cite{LouNP}. Here (see Fig. 2), a probe laser detects
the $\hat{z}$-component of electrically-injected spins via the
Hanle-Kerr effect, $\theta_K$ versus $B_x$, at a point $\sim$10
$\mu$m to the right of contact \#4. These $\theta_K (B_x)$ curves
exhibit similar ``local Hanle" lineshapes as seen in Fig. 1(d).
Extrema occur at $\pm B_x$ for which the spins have on average
precessed $\pm$90$^\circ$ (from $\pm \hat{y}$ to $\pm \hat{z}$) at
the detection point. A series of curves are shown in Fig. 2(b) at
various interface biases, $V_{int}$, across contact 4. Their
peak-to-peak amplitude $\Delta \theta_K$ is proportional to the
difference between injected majority- and minority-spin densities,
which necessarily scales directly with the difference between
majority- and minority-spin tunneling currents: $\Delta \theta_K
\propto n_\uparrow - n_\downarrow \propto j_\uparrow -
j_\downarrow$. Fig. 2(c) shows $\Delta \theta_K (V_{int})$ and the
current-voltage trace for contact 4. Normalizing by the total
current $j$ reveals the relative polarization of the interface
tunneling current, $P_j$$\propto$$(j_\uparrow - j_\downarrow)/j$
[see Fig. 2(d)]. $P_j$ depends strongly on $V_{int}$, showing a
maximum (majority spin) polarization near +50~mV, and a sign
inversion at -90~mV forward bias. All contacts on sample A and on
other devices from this wafer show the same bias dependence of
$P_j$.

\begin{figure}[tbp]
\includegraphics[width=.45\textwidth]{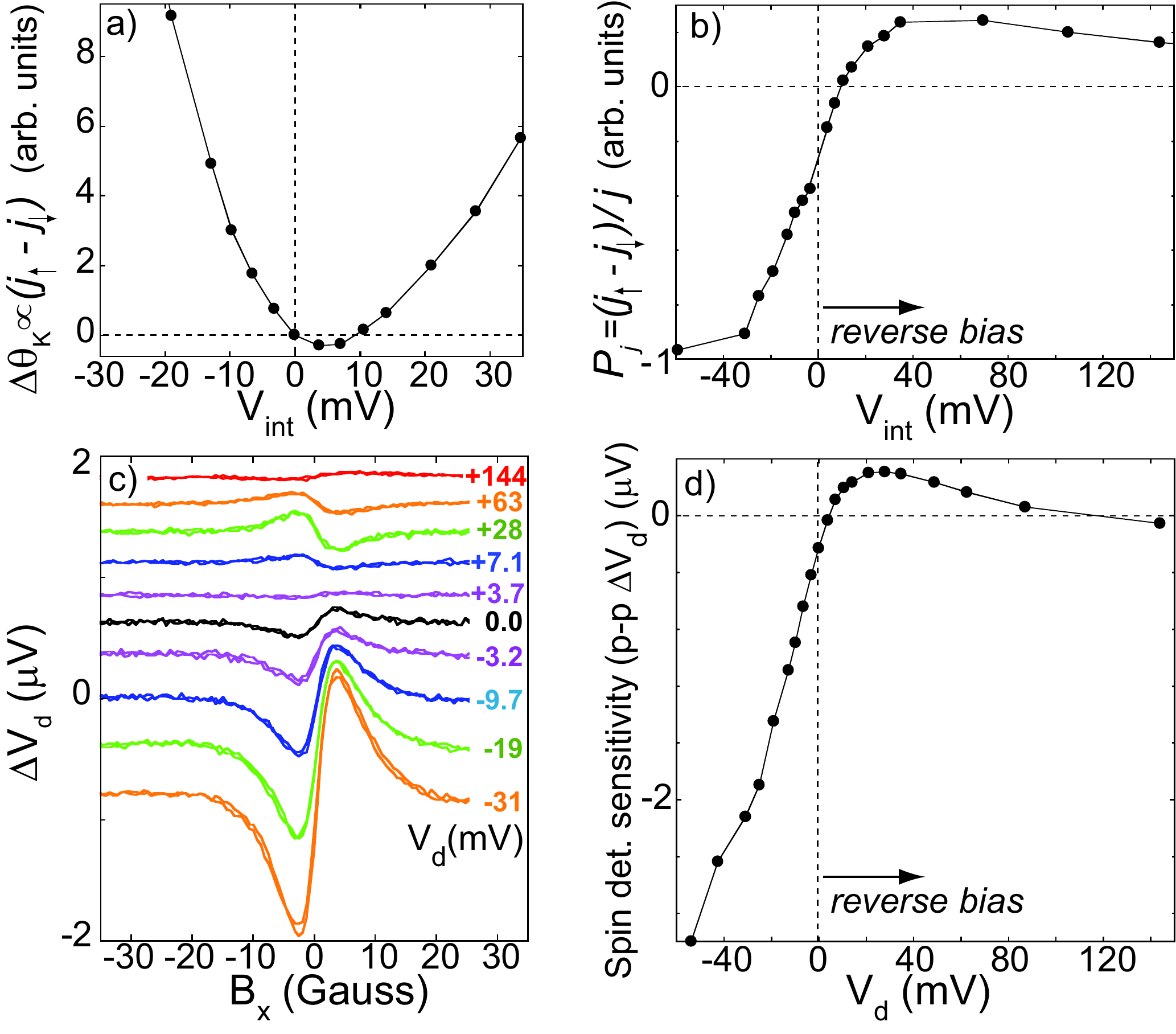}
\caption{(Color online) Spin injection/detection data from sample B at 10~K. a) The
electrically-injected spin density, measured by the Kerr effect
($\Delta \theta_K$ vs. $V_{int}$). b) Polarization of the tunneling
current $P_j$, which now inverts under \emph{reverse} bias, in
contrast to sample A. c) Raw spin detection data, $\Delta V_d (B_x)$
at various detector biases $V_{d}$ (curves offset). d) The
spin-detection sensitivity $\Delta V_d$ versus $V_{d}$.}
\label{fig3}
\end{figure}

Comparing sample A's spin-detection sensitivity $\Delta V_d$ with
$P_j$ [Figs. 1(e) and 2(d)], general trends can now be identified:
Both exhibit a positive maximum near zero bias, and the inversion of
$\Delta V_d$ under small forward bias can now be understood by the
similar inversion of $P_j$. However, a key finding is that striking
differences exist at large interface bias: Compared to $P_j$,
$|\Delta V_d|$ is markedly \emph{enhanced} under large forward bias,
but is \emph{suppressed} under large reverse bias ($\Delta V_d$ also
inverts sign again at +130 mV, which is unexpected and does not
follow from $P_j$). Thus, spin-detection sensitivities in these
Fe/GaAs structures do \emph{not} simply track $P_j$ as might be
expected from reciprocity arguments, and their strong deviation at
larger biases suggests the important and nontrivial role of electric
fields in the GaAs, discussed below.

Fig. 3 shows corroborating data from sample B. In contrast to sample
A, injection studies [Figs. 3(a,b)] show that $P_j$ is now dominated
by \emph{minority} spins at zero bias and that $P_j$ now inverts at
finite \emph{reverse} bias ($\sim$+10 mV). These differences likely
originate in the microscopic details of this Fe/GaAs interface,
which are not yet fully understood \cite{DeryPRL, ChantisPRL}.
Nonetheless, the sensitivity of sample B's spin detectors [Figs.
3(c,d)] can be correlated with $P_j$. At zero bias, the remotely
injected spins now generate small \emph{negative} voltage changes at
detection electrodes ($\Delta V_d \simeq$-0.2 $\mu$V), which now
invert sign at small reverse detector bias, similar to $P_j$.
However, like sample A, $\Delta V_d$ diverges markedly from $P_j$ at
large bias: $\Delta V_d$ is suppressed at large reverse bias, but is
greatly enhanced -- over tenfold -- with only -40 mV forward bias,
transforming a poor zero-bias spin detector into a much more
sensitive spin detection device.

\begin{figure}[tbp]
\includegraphics[width=.45\textwidth]{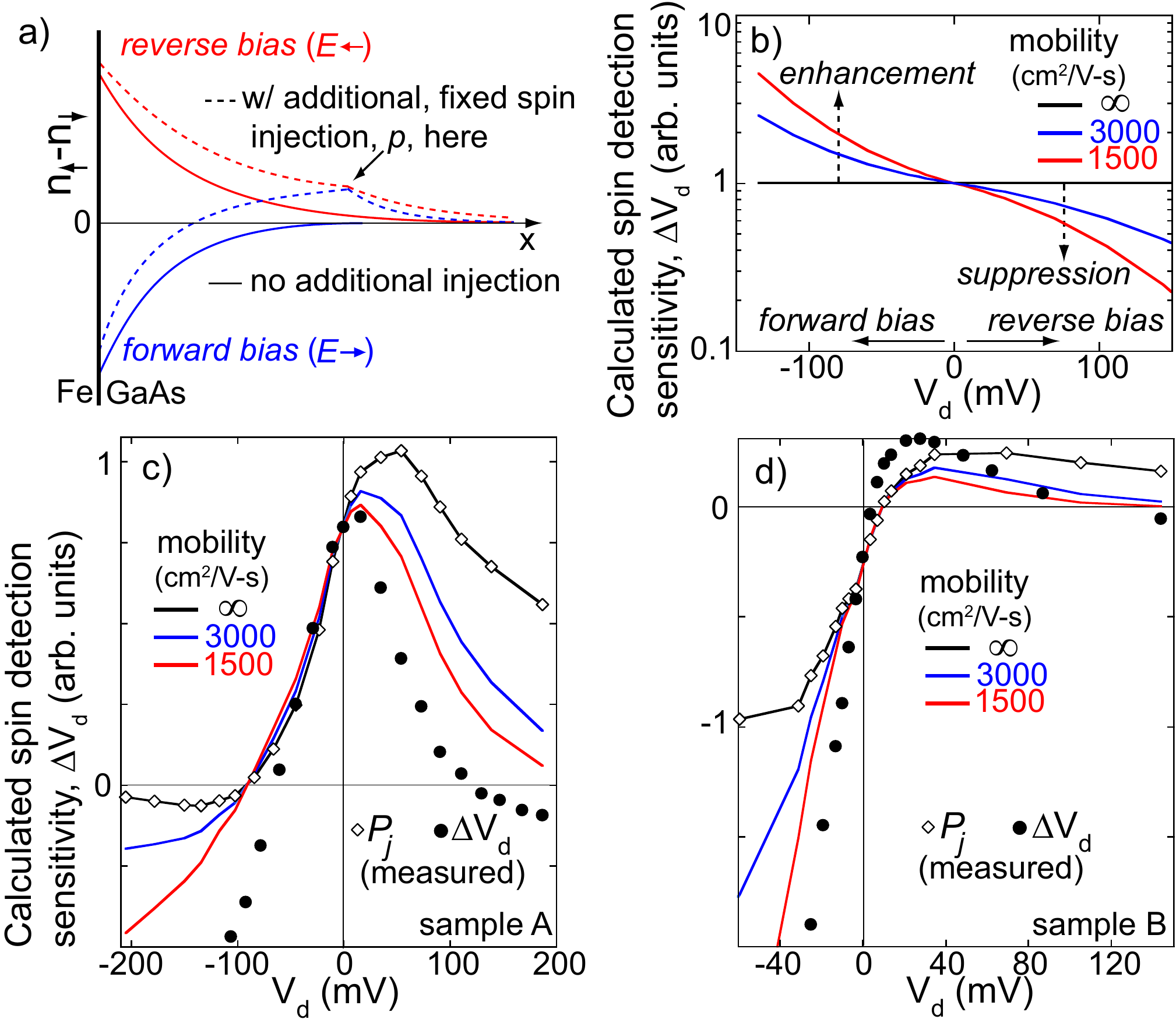}
\caption{(Color online) a) Schematic of $n_\uparrow - n_\downarrow$ near a Fe/GaAs
spin detector under forward/reverse bias $V_d$ (solid lines). Both
$n_\uparrow - n_\downarrow$ and its gradient are modified (dashed
lines) by a remotely-injected polarization, $p$. b) Calculated
sensitivity of a spin detector having \emph{constant} $P_j$.
Enhancement and suppression are due to $E$-fields in the GaAs. c,d)
Calculated $\Delta V_d$ for samples A and B, for three electron
mobilities $\mu$, using measured $P_j$ (open symbols) as inputs and
$p$ injected 20 $\mu$m away.} \label{fig4}
\end{figure}

The enhancement and suppression of $\Delta V_d$ can be understood
within a 1-D model of spin transport in the semiconductor and its
non-trivial influence on spin detection \cite{Athanasios}. As Fig.
4(a) depicts, a unit polarization $p$ generated in the channel
modifies the spin density difference $n_\uparrow- n_\downarrow$ and
its gradient at a nearby Fe/GaAs detector interface (in the figure,
both increase). At a fixed interface current $j$, these
modifications necessarily change $P_j$ if
$G_\uparrow$$\neq$$G_\downarrow$. From Eq. (1), one can derive the
induced change in detector voltage: $\Delta V_d \propto k \frac{j
\partial P_j }{\partial p}$ = $k \left[ D\frac{\partial }{\partial
p} \left[ \frac{\partial \left( n_ \uparrow - n_\downarrow
\right)}{\partial x} \right] + \mu E \frac{\partial }{\partial p}
\left( n_ \uparrow - n_ \downarrow \right)\right]$, where
$k$=$\frac{e}{4} \left( \frac{1}{G_ \uparrow } -\frac{1}{G_
\downarrow}\right)$, $D$ and $\mu$ are the electron diffusion
constant and mobility, $E$ is the electric field in the
semiconductor, and $n_ \uparrow - n_\downarrow$ and its gradient are
evaluated at the interface. Using the spin transport model of
\cite{Smith} and the full spin-drift-diffusion equations, one finds
\cite{Athanasios} that $\Delta V_d$ can deviate markedly from $P_j$
because: i) trivially, $E$ can drift $p$ away from (towards) the
detector in reverse (forward) bias, while ii) \emph{at the interface
itself}, the diffusion and drift terms oppose each other in reverse
bias ($E$$<$0) but add in forward bias ($E$$>$0). The importance of
(ii) is clearly seen in Fig. 1(e), where nearly identical detection
sensitivities are measured even when $p$ is injected \emph{in} the
current path. Note that $|E|$$<$10~V/cm -- rather modest -- in these
studies.

These enhancement and suppression effects are normally superimposed
on the bias dependence of $P_j$, but are disentangled in Fig. 4(b)
by showing $\Delta V_d$ calculated for an idealized (constant)
$P_j$. Crucially, detection sensitivities are enhanced or suppressed
relative to $P_j$ due to $E$, which depends sensitively on the
channel conductivity. For ``metallic-like" channels, $\Delta V_d$
tracks $P_j$ exactly, because $E$$\sim$0. Notably, this model
suggests that $\Delta V_d$ can also be controlled (independent of
$P_j$) by engineering the conductivity of semiconductor devices.

We explicitly calculate $\Delta V_d$ for samples A and B using
$G_{\uparrow, \downarrow}$ determined from experimental $P_j$ and
$j(V_{int})$ data. Electron densities and spin lifetimes are
measured independently. The results [Figs. 4(c,d)] are scaled so
that $\Delta V_d \equiv P_j$ at zero bias. For realistic
\emph{n}:GaAs mobilities, the calculated $|\Delta V_d|$ drops
(increases) much more rapidly than $P_j$ under large reverse
(forward) bias, in good overall agreement with measured $\Delta V_d$
values. (The inversion of $\Delta V_d$ at large reverse bias is not
reproduced in this 1-D model because drift and diffusion terms
cancel exactly, suppressing $\Delta V_d$ monotonically to zero
\cite{Athanasios}. No such restriction exists in 2-D or 3-D
geometries, which permit inhomogeneous $E$ fields.)

Finally, these results suggest that it should be possible to tune
and enhance the detection of \emph{electrically}-injected spins.
This is demonstrated in Fig. 5, where non-local lateral spin-valve
studies \cite{LouNP} were performed on a device having $P_j$ similar
to sample B. Fig. 5(a) shows the spin-valve signal versus detector
bias, when the \emph{source} electrode is forward- or
reverse-biased. The dependence on $V_d$ is the same, confirming that
a detector's sensitivity can be optimized independently of an
injector's biasing. Fig. 5(b) shows raw spin-valve data with
$V_d$=-150, 0, and +82 mV, explicitly showing that spin-detection
sensitivities are freely tunable in both sign and magnitude in
all-electrical devices.

\begin{figure}[tbp]
\includegraphics[width=.45\textwidth]{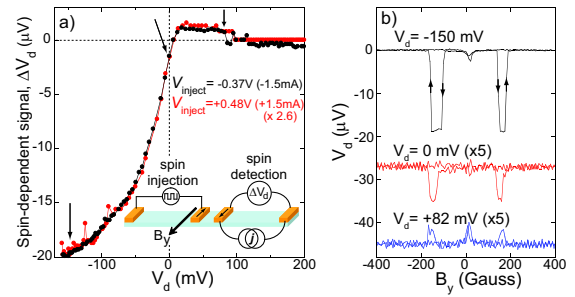}
\caption{(Color online) a) Inset: All-electrical lateral spin-valve setup. Black
(red) points show the spin-valve signal $\Delta V_d$ versus detector
bias $V_d$ for a forward (reverse) biased spin \emph{injector}. b)
Raw spin-valve data at three detector biases [arrows in (a)], using
a fixed -1.5 mA injector bias (curves offset). Note sign switching
and ten-fold enhancement of the detected signal.} \label{fig5}
\end{figure}

In these studies, both interfacial biases and electric fields play
essential roles. Because semiconductors can support large electric
fields, it is possible to bias each element -- source, channel, and
detector -- independently in multiterminal devices. Controlling spin
transport through both interfacial and bulk band structures
represents a unique capability of ferromagnet-semiconductor devices
that is only beginning to be explored. The authors acknowledge
support from the Los Alamos LDRD program, ONR, NSF MRSEC, NNIN and
IGERT programs.


\begin{references}

%\bibitem{DattaDas}Datta, S. \& Das, B. Electronic analog of the
%electro-optic modulator, \emph{Appl. Phys. Lett.} \textbf{56}, 665
%(1990).

\bibitem{Zutic}I. \v{Z}uti\'{c}, J. Fabian, and S. Das Sarma, Rev. Mod. Phys. \textbf{76},
323 (2004).%Spintronics: Fundamentals and applications.

\bibitem{Jansen}R. Jansen and B. C. Min, Phys. Rev. Lett. \textbf{99}, 246604 (2007).
%Detection of a spin accumulation in nondegenerate semiconductors.

\bibitem{Johnson2}P. R. Hammar and M. Johnson, Phys. Rev. Lett. \textbf{88}, 066806 (2002).
%Detection of spin-polarized electrons injected into a two-dimensional electron gas.

\bibitem{Bhatta}D. Saha, M. Holub, and P. Bhattacharya, Appl. Phys. Lett. \textbf{91}, 072513 (2007).
% Amplification of spin-current polarization.

\bibitem{LouNP}X. Lou, C. Adelmann, S. A. Crooker, E. S. Garlid, J. Zhang, K. S. M. Reddy, S. D. Flexner, C. J. Palmstr{\o}m, and P. A. Crowell, Nat. Phys. \textbf{3}, 197 (2007).
%Electrical detection of spin transport in lateral ferromagnet-semiconductor devices.

\bibitem{Huang}I. Appelbaum, B. Huang, and D. J. Monsma, Nature \textbf{447}, 295 (2007).
%Electronic measurement and control of spin transport in silicon.

\bibitem{vantErve}O. M. J. van 't Erve, A. T. Hanbicki, M. Holub, C. H. Li, C. Awo-Affouda, P. E. Thompson, and B. T. Jonker, Appl. Phys. Lett. \textbf{91}, 212109 (2007).
%Electrical injection and detection of spin-polarized carriers in silicon in a lateral transport geometry.

\bibitem{Tombros}N. Tombros, C. Jozsa, M. Popinciuc, H. T. Jonkman, and B. J. van Wees, Nature \textbf{448}, 571 (2007). %Electronic spin transport and spin precession in single graphene layers at room temperature.

\bibitem{Fert}M. Tran, H. Jaffr\`{e}s, C. Deranlot, J.-M. George, A. Fert, A. Miard, and A. Lema\^{\i}tre, Phys. Rev. Lett. \textbf{102}, 036601 (2009).

%\bibitem{Johnson1}M. Johnson and R. H. Silsbee, Phys. Rev. Lett. \textbf{55}, 1790 (1985).
%Interfacial charge-spin coupling: Injection and detection of spin magnetization in metals.

%\bibitem{Jedema} F. J. Jedema \emph{et al.}, Nature \textbf{416}, 713 (2002).
%Electrical detection of spin precession in a metallic mesoscopic spin valve.

%\bibitem{ValenzuelaPRL}S. O. Valenzuela \emph{et al.}, Phys. Rev. Lett. \textbf{94}, 196601 (2005). %Spin polarized tunneling at finite bias.

\bibitem{MJcomment}M. Johnson and R. H. Silsbee, Phys. Rev. Lett. \textbf{60}, 377 (1988).

\bibitem{Rashba}E. I. Rashba, Phys. Rev. B \textbf{62}, R16267 (2000) %Theory of electrical spin injection: Tunnel contacts as a solution of the conductivity mismatch problem.

\bibitem{Smith} D. L. Smith and R. N. Silver, Phys. Rev. B \textbf{64},045323 (2001).
%Electrical spin injection into semiconductors.

\bibitem{Moser}J. Moser, M. Zenger, C. Gerl, D. Schuh, R. Meier, P. Chen, G. Bayreuther, W. Wegscheider, D. Weiss, C.-H Lai, R.-T Huang, M. Kosuth, and H. Ebert, Appl. Phys. Lett. \textbf{89}, 162106 (2006). %Bias dependent inversion of tunneling magnetoresistance in Fe/GaAs/Fe tunnel junctions.

\bibitem{DeryPRL}H. Dery and L. J. Sham, Phys. Rev. Lett. \textbf{98}, 046602 (2007).
%Spin extraction theory and its relevance to spintronics.

\bibitem{ChantisPRL}A. N. Chantis, K. D. Belashchenko, D. L. Smith, E. Y. Tsymbal, M. van Schilfgaarde, and R. C. Albers, Phys. Rev. Lett. \textbf{99}, 196603 (2007). %Reversal of spin polarization in Fe/GaAs(001) driven by resonant surface states: First-principles calculations.

\bibitem{YuFlatte}Z. G. Yu and M. E. Flatt\'{e}, Phys. Rev. B \textbf{66}, 201202(R) (2002). %Electric-field dependent spin diffusion and spin injection into semiconductors.

\bibitem{Schmidt}G. Schmidt, C. Gould, P. Grabs, A. M. Lunde, G. Richter, A. Slobodskyy, and L. W. Molenkamp, Phys. Rev. Lett. \textbf{92}, 226602 (2004). %Spin injection in the nonlinear regime: Band bending effects.

\bibitem{Hanbicki}A. T. Hanbicki, O. M. J. van 't Erve, R. Magno, G. Kioseoglou, C. H. Li, B. T. Jonker, G. Itskos, R. Mallory, M. Yasar, and A. Petrou, Appl. Phys. Lett. \textbf{82}, 4092 (2003).
%Analysis of the transport process providing spin injection through an Fe/AlGaAs Schottky barrier.

\bibitem{Stephens}J. Stephens, J. Berezovsky, J. P. McGuire, L. J. Sham, A. C. Gossard, and D. D. Awschalom, Phys. Rev. Lett. \textbf{93}, 097602 (2004). %Spin accumulation in forward-biased MnAs/GaAs Schottky diodes.

\bibitem{CrookerScience}S. A. Crooker, M. Furis, X. Lou, C. Adelmann, D. L. Smith, C. J. Palmstr{\o}m, and P. A. Crowell, Science \textbf{309}, 2191 (2005). %Imaging spin transport in lateral ferromagnet/semiconductor structures.

\bibitem{FurisNJP}M. Furis, D. L. Smith, S. Kos, E. S. Garlid, K. S. M. Reddy, C. J. Palmstr{\o}m, P. A. Crowell, and S. A. Crooker, New J. Phys. \textbf{9}, 347 (2007).
%Local Hanle-effect studies of spin drift and diffusion in \emph{n}:GaAs epilayers and spin-transport devices.

\bibitem{Kotissek}P. Kotissek, M. Bailleul, M. Sperl, A. Spitzer, D. Schuh, W. Wegscheider, C. H. Back, and G. Bayreuther, Nat. Phys. \textbf{3}, 872 (2007).

\bibitem{Athanasios}A. N. Chantis and D. L. Smith, Phys. Rev. B \textbf{78}, 235317
(2008).

\end{references}
\end{document}